%% file: narr.tex
\documentclass[onecolumn]{article}
\usepackage{narr}

\usepackage{amssymb,epsfig,graphicx,amssymb,subeqnarray, multirow, pstricks,colortbl, bm}
\usepackage{units,tabularx,array,booktabs,booktabs,algorithm, algorithmic}
\usepackage[cmex10]{amsmath}
\usepackage[tight,footnotesize]{subfigure}  
\usepackage{natbib,apalike}
\include{macpap2}

\usepackage{paralist}

\usepackage{hyperref}
\usepackage{soul}
\usepackage{float}
\usepackage[font={small}]{caption}

\definecolor{Gray}{gray}{0.9}
\newcolumntype{g}{>{\columncolor{Gray}}c}
\renewcommand{\cite}{\citep} 

\begin{document}

\title{A predictive coding account of OCD}

\author{P.J. Moore${}^\dag$ }

\address{
\dag Mathematical Institute, Oxford University.\\
}

\maketitle

\abstract
This paper presents a predictive coding account of obsessive-compulsive disorder (OCD). We extend the predictive coding model to include the concept of a `formal narrative', or temporal sequence of cognitive states inferred from sense data.  We propose that human cognition uses a hierarchy of narratives to predict changes in the natural and social environment.   Each layer in the hierarchy represents a distinct view of the world, but it also contributes to a global unitary perspective.  We suggest that the global perspective remains intact in OCD but there is a dysfunction at a sub-linguistic level of cognition.  The consequent failure of recognition is experienced as the external world being `not just right', and its automatic correction is felt as compulsion.  A wide variety of symptoms and some neuropsychological findings are thus explained by a single dysfunction.   We conclude that the model provides a deeper explanation for behavioural observations than current models, and that it has potential for further development for application to neuropsychological data.
\endabstract

\keywords{obsessive-compulsive disorder, ocd, bayesian inference, predictive coding, generative models, prospection, mental imagery}

%
\section{Introduction}
This paper presents an account of obsessive-compulsive disorder (OCD) based on a predictive coding model of cognition.  An explanatory framework is presented and it is used to construct an account of OCD with a hypothesis for the underlying dysfunction. The paper is organised as follows:  this section gives the context of the research by describing symptoms and theories of the disorder. Next, we describe the explanatory framework in detail and give an account of OCD with explanations for empirical results and symptoms. We suggest experimental work to test the hypothesis and establish its association with OCD. Finally we consider how the account compares with alternative models for OCD in terms of parsimony and fit to known data.
%
%
\subsection{Observations}
The symptoms of OCD are obsessions and compulsions.  Obsessions are recurrent and persistent thoughts, impulses, or images which are experienced as intrusive and inappropriate, and they cause marked anxiety or distress \cite{DSM}.  Compulsions are repetitive behaviours or mental acts that the person feels driven to perform.  Some examples of symptoms are: repeatedly checking gas taps or washing hands; fearing having knocked someone down while driving; feeling an impulse to shout obscenities during a church service \cite[p196]{miller1988}.   More generally the focus of obsessions can be aggression, contamination and symmetry among others, while compulsions are often centred on checking, ordering, cleaning and hoarding \cite{DSM}.  The categories are related: for example obsessions about aggression, religious or sexual themes are associated with checking compulsions \cite{summerfeldt1999,leckman97}.  Some patients exhibit common cognitive traits or beliefs: 1) Responsibility and threat estimation, 2) Perfectionism and intolerance for uncertainty, and 3) Importance and control of thoughts \cite{occwg03_1,occwg03_2}, while other patients do not exhibit dysfunctional beliefs \cite{antony02,taylor06}.  Some patients believe that the intrusive thoughts can influence events in the world, a phenomenon known as `thought--action fusion' \cite{rachman1993,shafran2004}. 
\nind
A meta analysis of 113 studies found reduced performance in people with OCD compared with healthy individuals across most neuropsychological domains \cite{abramovitch2013}.  Those with OCD were found to score worse than controls on non-verbal memory tasks, with a much smaller effect size for verbal memory tasks, although there are questions about the involvement of impaired executive functioning and the effect of medication on the non-verbal memory tasks \cite{abramovitch2013,savage1999,shin2010}. Neuroimaging studies of OCD have found differences between patients with OCD and controls in the orbital gyrus and the head of the caudate nucleus \cite{saxena2001,whiteside2004}.  A review of evidence from both neuroimaging and neuropsychological studies is given in \cite{menzies2008}.


%
\subsection{Theories}
Some theoretical approaches to OCD emphasise the negative appraisal of intrusive thoughts, where such appraisals are engendered by dysfunctional beliefs.  This cognitive-behavioural account of OCD begins by noting that intrusive thoughts and images, similar in content to clinical obsessions, occur generally in the population \cite{julien2007,gibbs1996,rachman1978}. The hypothesis is that in OCD such intrusions develop into obsessions when they are appraised as personally important, highly unacceptable or immoral, or as posing a threat for which the individual is personally responsible \cite{abramowitz2006,abramowitz2007}.  Compulsions arise from an attempt to remove intrusions and prevent their harmful consequences, and these actions serve to increase the frequency of intrusions by acting as a reminder of their content.  This account of OCD was originated by several researchers \cite{mcfall1979,rachman1997,salkovskis1985}.  Beliefs, appraisals and symptoms have some correlation \cite{abramowitz2006,abramowitz2007}, and the pattern of this association has been adduced as evidence for the cognitive-behavioural model of obsessive-compulsive disorder \cite{abramowitz2007}. Some objections to the approach were articulated by Jakes \cite{jakes1989a,salkovskis1989,jakes1989b}, and a overview of criticisms within a wider context was given by Jakes \cite{jakes}. A more recent critique of the significance of dysfunctional appraisals and the cognitive-behavioural account was given by Cougle and Lee \cite{cougle2014}.  Some researchers have identified dimensions of harm avoidance and incompleteness as more fundamental motives that contribute to compulsive behaviour \cite{summerfeldt2004}, where incompleteness is defined as an internal state of imperfection or `not just right' experience \cite{coles2003}.  
\nind
Other theories of OCD have implicated problems with memory \cite{muller2005}, while another research strand focuses on habit \cite{graybiel2000,gillan2014}.  Some researchers have emphasised the importance of biological factors: a brief theoretical overview of biological and other models is provided in \citet{abramowitz2009}.  Wise and Rapoport suggested that the disorder arises from a dysfunction of the basal ganglia \cite{wise1989,rapoport1990}.   OCD can occur in childhood, associated with streptococcal infections, as part of the paediatric autoimmune neuropsychiatric disorders associated with streptococcal infections (PANDAS) syndrome \cite{swedo2002,leonard2001}.  The related hypothesis is that OCD (and tic disorders) arise from post-streptococcal immunity.  A neuropsychological model, relevant to the current study,  explains OCD as a disturbance of security motivation \cite{szechtman2004}.  In that account the symptoms of OCD stem from an inability to generate a `feeling of knowing' that normally terminates the expression of a security motivation system. Some criticisms of this model were given in \citet{taylor2005} and a response was given in \citet{woody2005}.  The context of the current study is that there are difficulties with the dominant theoretical approach \cite{jakes,cougle2014} and there remains as yet no definitive account of OCD cf. \citet[p88]{taylor06} and \citet[p165]{jakes}.  
%
\subsection{Predictive coding}
The symptoms of OCD suggest no obvious common cause and the disorder is heterogeneous, possibly with different subtypes \cite{mckay2004}.  Some psychological models for OCD do not go far beyond a description of the symptoms or traits of the patient \cite[p42]{jakes} and so they have weak explanatory power.  Biological models explain observed abnormalities in, for example imaging data, but they do not provide the level of description needed to explain symptoms.  In explaining mental illnesses, the different levels of explanation -- qualitative psychological explanations of symptoms, and formal electrophysical models at the neural level -- have tended to be disjoint.  There is some work that attempts to close this `explanatory gap', in particular the \emph{Bayesian brain} hypothesis \cite{knill,fristonhist} which has been applied to both neuroscientific observations \cite{fristoncort} and symptoms of mental illness \cite{fletcher,fristonhal}.  Under this approach the brain is hypothesised to maintain probabilistic models of the environment and update the models using Bayesian inference \cite{knill}.  More specifically, the brain minimises the discrepancy between sensory input and the predictions made by an internal model, and in this way it implements Bayesian inference.  The explanatory power of this \emph{predictive coding} model is explored in depth in \citet{hohwy}.  The Bayesian brain concept is usually traced to  Helmholtz's theory of perception \cite{helm,southall1925} in which stimuli are seen as insufficient to generate percepts without prior information enabling unconscious inference.  O'Callaghan describes the development of perceptual theory from Helmholtz to a contemporary understanding \cite[p78]{frankish2012}, and Friston provides a short history of the Bayesian brain idea from a neuroscientific point of view \cite{fristonhist}.  
\nind
Two relevant applications of predictive coding to psychopathology are as follows.  First, Fletcher and Frith use a hierarchical model of brain function to explain hallucinations and delusions in schizophrenia \cite{fletcher}. In a hierarchical model, prediction errors at a low level are passed up the hierarchy until they are resolved or `explained away' by a higher level.  Fletcher and Frith postulate a failure in inference which leads to improper integration of new evidence and a resulting prediction error.  Second, Corlett \etal use a predictive coding model to explain delusions \cite{corlett}.  Their hypothesis is that aberrant prediction error leads to learning failure and this in turn leads to delusions and perceptual aberrations. The approach taken in the current paper also uses a hierarchical model of cognition and proposes a hypothesis for its dysfunction, in this case to explain the symptoms of OCD.

%
\section{A hierarchical narrative framework}

%
\subsection{Hierarchical inference}
We first explain hierarchical Bayesian inference as a model for human cognition.  We define an \emph{external state} $X$ as a cause or condition in the environment that fully determines the \emph{observations} $U$.  
%

\begin{figure}[htp]
\centering
\includegraphics[trim = 0mm 0mm 0mm 0mm, width=12cm]{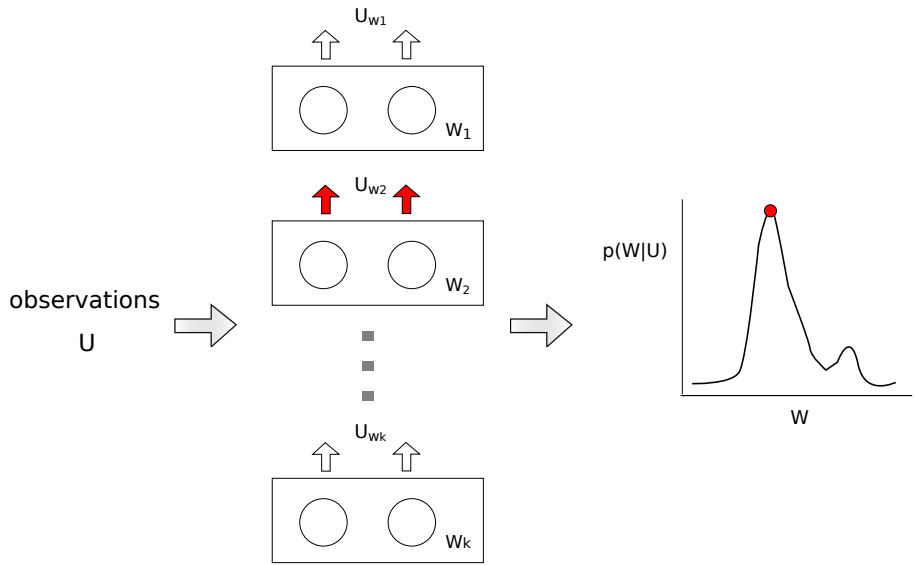} 
\caption[Recognition as cognitive resonance]{Recognition as cognitive resonance.  The rectangular boxes represent internal state models  $W_1$ \dots $W_k$ each of which can generate simulated observations $U_{W1}$ \dots $U_{Wk}$.  The internal states $W$ might be visual features such as edges, or higher level percepts such as a human face.   Perception is the process of inferring the model that best explains current observations $U$ by using Bayes' rule to find $p(W|U)$ from $p(U|W)$.  In the figure, $W_2$ is shown as resonating strongly with the observations. The process results in the graph $p(W|U)$ which has a peak at $W_2$ (right hand side).  A familiar pattern will exhibit resonance, while closely related patterns also resonate to some extent.  For example, an infant will recognise any human face, but will show particular affinity for the mother's face.}
\label{fig:perception}
\end{figure}
\begin{figure}[!htp]
\centering
\includegraphics[trim = 0mm 0mm 0mm 0mm, width=8cm]{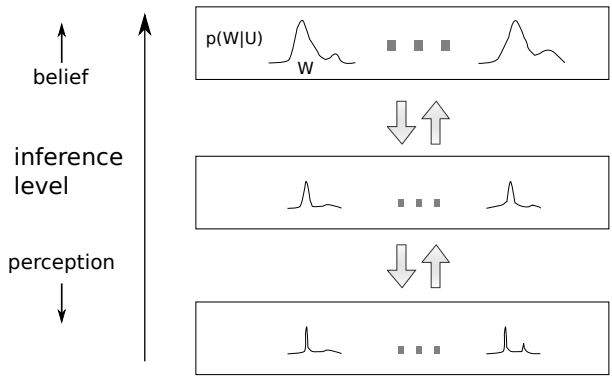} 
\caption[Cognition as multi-level inference of states]{A pictorial representation of cognition as multi-level inference of inferred states.   The graphs represent the conditional distribution $p(W|U)$ which has a peak at the most likely state.  Inferences from the lowest level may consist of features, such as edges in the visual scene, which become observations for the next level in the hierarchy. High level percepts needed for beliefs, thoughts and imagination are represented at the top of the hierarchy.  The perception--belief spectrum was suggested by Corlett to be relevant to delusions and hallucinations in schizophrenia \cite{corlett}.} 
\label{fig:hierarchy}
\end{figure}

We define an \emph{internal state} $W$ as the cognitive representation of an external state based on the observations.  The task for the cognitive apparatus is to find the distribution of internal states conditioned on observations, $p(W|U)$ in order to model the external state. Figure \ref{fig:perception} illustrates the process.  The probability of each state $W_1$ \dots $W_k$ individually generating the observations $U$ is determined and the states that are most likely to have generated the observations $U$ are preferred.  The result is shown on the right hand side as a graph of $p(W|U)$ against the internal states $W$.  For a known pattern generated by a familiar state, the graph will show resonance for the corresponding model.
\nind
Higher level inference is accomplished by using a hierarchy in which states inferred by one level are used as observations for the next, as shown in Figure \ref{fig:hierarchy}. Observations arrive from sensory neurons as spike trains which are processed into features, such as edges in a visual scene.  The features themselves become observations for a second level of inference which in turn uses them to infer states that are meaningful to the next level up.  So higher levels attempt to predict or `explain away' lower levels, so that ultimately the sense data is explained by the internal cognitive model.

%

%

\subsection{Formal narratives}
The generative models used for inference can also generate artificial observations which have the same distribution as represented in the model.  We illustrate the process by using a generative model of language to create English sentences.  The probability of a sequence of words $W_k, (w_1, w_2, \dots w_k)$ can be expanded as,
\begin{align}
p(W_k) &= p(w_1)p(w_2|w_1)p(w_3|w_1,w_2) \dots p(w_k|w_1 \dots w_{k-1}) 
\intertext{We can approximate the terms in the expansion by limiting the history length to give a bigram word model, which assumes that the probability of each word is influenced only by its predecessor,}
p(W_k) &\approx p(w_1)p(w_2|w_1)p(w_3|w_2) \dots p(w_k|w_{k-1}) 
\end{align}
The model then comprises the frequencies of starting words and pairs of words found in the training data. To create a new sentence, we take a starting word $w_1$ and choose the next word randomly according to its distribution in the model, and continue until the desired length is reached.  Shannon \cite[p7]{shannon} gives an example of an artificial sentence derived from a similar model of word sequences,
\\ \\
\texttt{\phantom{THE} THE HEAD AND IN FRONTAL ATTACK ON AN ENGLISH WRITER THAT THE CHARACTER OF THIS POINT IS 
\\ \phantom{THE} THEREFORE ANOTHER METHOD FOR THE LETTERS THAT THE TIME OF WHO EVER TOLD THE PROBLEM FOR 
\\ \phantom{THE} AN UNEXPECTED.}
\\ \\
\comment{In the last example the first word \texttt{THE} is likely to be followed by a noun or adjective, in this case \texttt{HEAD}, and it in turn  determines the distribution of the next word.} The sentence as a whole approximates the distribution of the natural word order.  If a bigram model is used, the generated text is word salad.  Longer n-grams can be used, in which case the model comprises longer sequences of words. In this case the generated text becomes much more recognisably like English in its construction, but it usually has no coherent meaning.
\nind

\begin{figure}[htpb]
\centering
\includegraphics[trim = 0mm 0mm 0mm 0mm, width=7cm]{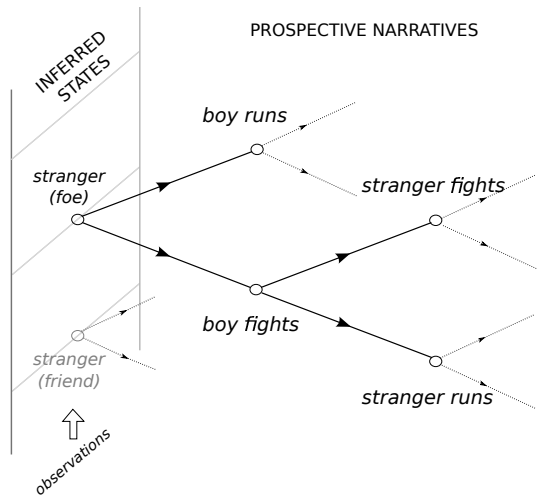} 
\caption[Example of prospective narratives]{Example of prospective narratives generated from a model of states.  The internal states which represent the stranger as a foe or a friend are inferred from observations.  These states are those that best `explain away' or predict the observations, which are themselves inferred states from lower in the cognitive hierarchy.  A network of prospective narratives is then generated, whose probabilities depend on past experience.  }
\label{fig:narratives_prosp}
\end{figure}

For modelling cognition we apply the same process of generation, but instead of words, we use the cognitive states $W$ as generated elements.  We call a sequence of generated states a \emph{formal narrative}.   To illustrate the idea we use the following situation from everyday life: a boy is walking home from school when he sees another boy in the distance.  He makes an assessment of whether the stranger is a friend or a foe, and then acts accordingly.  We denote the internal state representing the stranger as a foe, $W_{stranger-foe}$, and the internal state representing the boy running away, $W_{boy-runs}$, and so on.  Some prospective narratives representing this fight--or--flight scenario are,
\newline\newline
\indent\indent $N_1$ = \{$W_{stranger-foe}, \quad W_{boy-runs}$\} \\ 
\indent\indent $N_2$ = \{$W_{stranger-foe}, \quad  W_{boy-fights}, \quad  W_{stranger-fights}$\} \\
\indent\indent $N_3$ = \{$W_{stranger-foe}, \quad  W_{boy-fights}, \quad  W_{stranger-runs}$\} \\
\indent\indent $N_4$ = \quad \texttt{$\dots$} 
\newline\newline
The scenarios are illustrated in Figure \ref{fig:narratives_prosp} which shows two inferred states, representing the stranger as a friend and as a foe and their network of prospective narratives, 

%
\subsubsection{Hierarchies of narratives}
\begin{figure}[!htp]
\centering
\includegraphics[trim = 0mm 0mm 0mm 0mm, width=13cm]{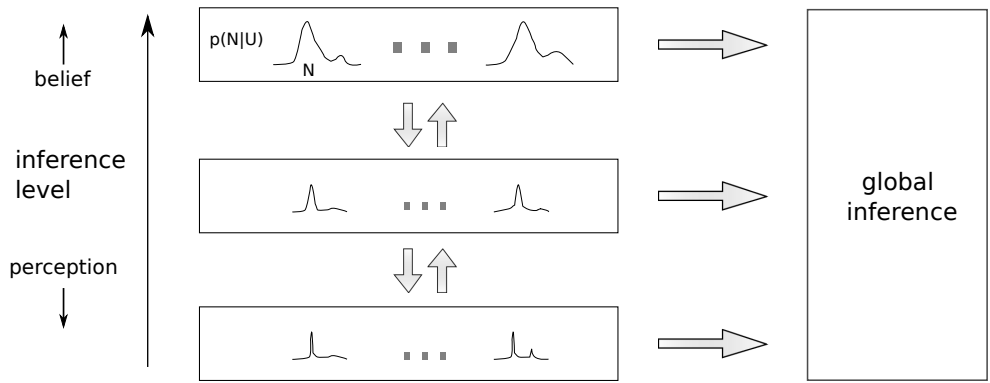} 
\caption[]{A hierarchical narrative model of cognition.  The graphs represent the conditional distribution $p(N|U)$ which has a peak at the most probable narrative given the sense data.  Each layer in the hierarchy represents a different perspective on the external world using a set of narratives.  All the layers contribute to global inference, shown on the right, which represents unitary perception.} 
\label{fig:hierarchy_narr}
\end{figure}

The notation for states in the example, such as  $W_{stranger-foe}$, uses English because the states all lie at the linguistic level of cognition.  The situation can also be represented using narratives in the more colloquial sense.  So the formal narrative $N_3$ can be expressed retrospectively as, \emph{``I saw a unfriendly stranger, I fought him and he ran away''}.  The substates from which the state $W_{stranger-foe}$ is inferred can not so easily be expressed in a natural language, but there is evidence that more primitive perceptions are also the outcome of a generative process.  For example, the visual hallucinations that occur in Charles Bonnet Syndrome have been proposed as evidence for a generative model of vision \cite{reichert2013}.  This point suggests that generated narratives could occur at levels of inference below those that can be expressed in a natural language, the `sub-linguistic' level of cognition.  In Figure \ref{fig:hierarchy_narr} the hierarchy is a generalisation of that in Figure \ref{fig:hierarchy} to allow inference from narratives, where a narrative is either a single inferred state or a sequence of inferred states.  This model has some intuitive appeal because it is usual to apply inference to narratives in everyday life.   For example someone hearing $N_3$ as a retrospective narrative might question if the boy really did have a fight, or if he just made up the story. 
%

%
\subsubsection{Cognition as a committee}
In the hierarchical model each layer holds a set of narratives which represents a `view' or perspective of the world.  For example, the lowest levels of inference may represent a visual scene composed of edges, while a high level might represent the behaviour of an adversary or friend.  Each layer contributes to the global perspective, which represents the unitary perception of common experience.  To illustrate this point, we use the metaphor of a multidisciplinary clinical team discussing a newly referred patient.  Each team member has their own view according to their expertise and knowledge about the patient: some members might have a wide expertise but not see the patient very often, while others have day-to-day contact. In an efficient team, the consensus is not simply the view of the most senior clinicians, but it is informed by and distinct from all the members' different views.  We argue that a similar process takes place with cognition: the unitary view is not simply the inference made by the highest level of cognition, but it is derived from all the layers.  
\nind
The philosophical study of unitary perception and consciousness has a long lineage, for example see \citet{bayneschol,bayne2010}. Support for a layered model of cognition is provided by instances when individuals do not believe what they see, as for example with optical illusions.  In these cases inference at a high level overrides inferences made by a subordinate level, so the individual has the experience of perception without its consequent belief.  The potential lack of coherence between levels is also relevant to OCD, where the individual recognises that the obsessional thoughts are a product of his or her own mind \cite{DSM,foa1995}.

%
\subsection{Generation of narratives}
There is a selective advantage to be derived from interpreting and predicting changes in the natural environment: an animal is more likely to reproduce if it can anticipate threats and rewards. One way of predicting changes in an undetermined environment is by the constant simulation of possible scenarios.  We propose that new, counterfactual narratives are created from the generative models used for inference.  The use of generative models for anticipation, rather than only for predicting current sensations, was suggested by \citet{friston2011}. A similar idea for employing generative models for prospection, retrospection and mental imagery was also put forward in \citet{thesis}.  Since narratives stem from percepts and unconscious inferences, as shown in Figure \ref{fig:narratives_prosp}, we can understand them as extending unitary perception into a manifold past and future\footnote{Simulation is relevant to both `mental time travel', the process of mentally projecting backwards or forwards in time \cite{suddendorf1997} and to some theories of theory of mind \cite{mitchell2009}.}.  Just as a movie passes through time using a series of 2-dimensional scenes, a single prospective narrative can run percepts forward in time, generating for each step a prospective scene.  However, whereas conscious perception is usually veridical in quality\comment{Sometimes percepts do not correspond with reality, for example in the case of hallucinations or perceptual illusions.}, prospective narratives are mostly counterfactual. 


%

\subsubsection{The security motivational system}
\label{sec:sms}
The risk of an unlikely adverse event is the focus of some symptoms of OCD, such as checking and fear of harm.   Szechtman and Woody suggest that a dysfunction in the security motivational system (SMS) underlies such symptoms, specifically an inability to generate the normal `feeling of knowing' that normally signals task completion \cite{szechtman2004}. The security motivational system refers to a `set of biologically based (hardwired), species-typical behaviours directed toward protection from danger of self and others' \cite[p113]{szechtman2004}.  In this paper we use the term to refer to the cognitive function that invokes those behaviours.  We suggest that the SMS simulates potential behaviour by creating a collection of narratives to guide the response. 
\nind
So we propose that, \emph{in response to a threat stimulus, a collection of narratives relating to that threat is generated.} These narratives would involve the highest levels of cognition but the response has to be fast, suggesting that the `view' of lower levels of cognition, using simpler inferred states, is also important.  This is a dual process approach which recognises two modes of cognition, a fast System 1 which is old in evolutionary terms, and a slower System 2 which is more recent \cite{evans2003}. We can compare the process with organisational security which performs penetration testing using scenarios to determine how attackers can gain access to sites or data.  Significantly, these scenarios include threats which arise from within the organisation either through action or inaction. In the human population there will be some variation in narratives generated for simulation: some individuals will generate many narratives having a low probability.  These unlikely narratives confer some protection in unpredictable circumstances, while a tendency towards more probable narratives makes for quick recognition. But if there is abnormal number of narratives having a low probability, we might expect there to be some dysfunction.

\section{A predictive coding account of OCD}
The notion of generated threat scenarios, including threats from within, speaks to `fear of harm' obsessions in OCD but it explains normal functioning rather than pathological obsessions.  The challenge for any account of OCD is to propose a dysfunction that explains symptoms and which is supported by other evidence.  For example, why are obsessions in OCD recurrent in nature, and why do they lead to marked distress? We begin by discussing the dysfunction, then we examine the results from Gillan \etalns's laboratory study of avoidance behaviour in OCD \cite{gillan2014}.   Next we apply the predictive coding account to specific symptoms of OCD, and finally to neuropsychological evidence.
%
\subsection{The nature of the dysfunction} 
\label{sec:dys}
\begin{figure}[htpb]
\centering
\includegraphics[trim = 0mm 0mm 0mm 0mm, width=8.5cm]{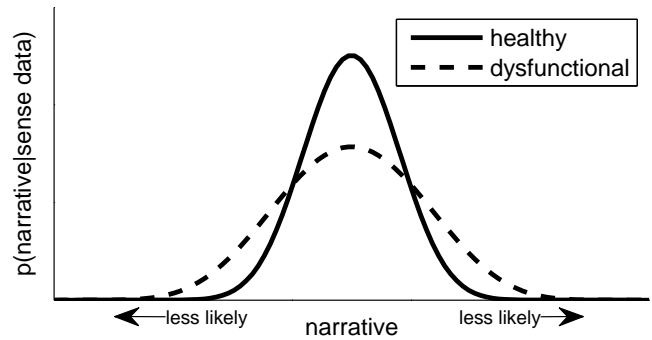} 
\caption[]{A representation of healthy and dysfunctional narrative distributions.  The x-axis lists sub-linguistic narratives which are generated in response to a threat scenario. The curves show the estimation of the probability of each narrative given the evidence from the senses.  The solid curve shows a healthy distribution with a peak at the most likely (veridical) narrative.  The dashed curve shows an unhealthy distribution which arises because there are too many unlikely scenarios.  There are two consequences to the dysfunction 1) unlikely events are collectively assigned too much probability mass, and 2) an abnormally low probability for veridical inference leading to impaired recognition by higher levels of inference.}
\label{fig:dysfn}
\end{figure}

\emph{Our hypothesis is that in OCD an abnormally high number of narratives having a low probability is generated.} In consequence the posterior distribution $p(N|U)$ is abnormally imprecise and inference, or recognition, becomes more uncertain. The dysfunction is located at a sub-linguistic level in the cognitive hierarchy, but the global System 2 inference remains intact: global inference is robust to failures of subordinate inference because it has evolved to manage uncertain sensory data. The dysfunction is illustrated in Figure \ref{fig:dysfn}.
\vspace{15mm}
%

When there is impaired recognition of a threat, an individual acts both to improve that inference and to respond to the threat.  For example, a driver slows down on approaching a traffic junction both to see the junction more clearly and to avoid a collision\footnote{The driver's behaviour can be seen as a manifestation of active inference, a unified account of both action and perception motivated by the free energy principle \cite{friston2010,fristonfree}.}. In the presence of a threat, the urge to correct a perceptual deficiency is compelling: consider the reaction of a driver whose vision suddenly becomes obscured.  The system for monitoring threats, the SMS, usually operates silently but in OCD attention returns repeatedly to a non-existent threat as the individual tries to improve precision.   This account explains why the compulsion to act is strong and why attention keeps returning to the perceived threat.  It also explains why individuals with OCD do not usually believe that a real threat exists: global inference is robust to deficiencies in the sensory data from an occluded scene, or to uncertainty in recognition. 
\nind
The robustness of global inference in the presence of uncertainty is illustrated by the following text, which has errors in most of the words,
\\ \\
\texttt{\phantom{THE}\phantom{THE}An illrtsaiuton of a haiirecrahcl Biseayan shceme for raednig. It sulohd be esay } \\
\texttt{\phantom{THE}\phantom{THE}to raed the sentncee `Jcak and Jlil wnet up the hlil'. You splmiy covenrt }  \\
\texttt{\phantom{THE}\phantom{THE}the shaeps itno ltteers, the lteters itno wdros and the wodrs itno sencestne.}
\\ \\
The original text was taken from an example in \citet{fletcher}.  Each word has been modified to keep the first and last letters in place while the remainder are shuffled.  Despite the word errors, the overall meaning can be inferred using prior knowledge and observed cues to predict the most likely word sequence.  Individual words can to be recognised quite accurately from their context, but not necessarily in isolation.  A consequence is that if we focus attention on one word in the hope of improving recognition, the effect is to impair it.  This phenomenon is relevant to obsessive re-checking, and to traits of perfectionism and intolerance of uncertainty in OCD.

%
\subsection{Gillan \etal -- Excessive avoidance habits}
Gillan \etalns's  study is important both for its results and for its method, which provides an experimental model for some kinds of compulsion \cite{gillan2014}.  It tested the possibility that excessive behavioural repetition in OCD is driven by a failure to learn about safety rather than being evidence for habit formation. The experiment compared a group of 25 individuals having OCD with a matched control group.  Participants were asked to avoid an electric shock by responding to a warning stimulus presented on the screen.  Two electrodes were used, one connected to each wrist and each delivering a shock at random intervals.  The shocks were preceded by a warning stimulus whose colour denoted either the right or left wrist as the target.  To avoid the shock, participants had to respond within 750ms using a corresponding left or right pedal switch.  The experimenters examined the effect of stimulus devaluation on goal--directed learning by disconnecting one of the electrodes in full view of the participant.  The task design consisted of four stages: a training session, a first devaluation test, an extended training session and a final devaluation test.  
\nind
In the first devaluation test, both the OCD group and the control group behaved in a similar way: they responded more to the valued stimulus which represented a real shock than to the devalued stimulus which did not.  So both OCD patients and control participants were capable of learning about safety.  After extended training, the OCD group showed greater avoidance of the devalued stimulus compared with the control group, but there were no measurable differences in contingency knowledge, explicit threat appraisal, or physiological arousal between the groups.  Those who responded to the devalued stimulus were asked to account for their action -- why did they continue to respond to the stimulus corresponding to the disconnected electrode?  The main categories of response were threat beliefs, `I thought it could still shock me', and accidental responses, `I lost concentration'.  The authors concluded that these findings supported an account of OCD involving habit formation.  How might Gillan \etalns's results be explained using the hierarchical narrative model?  We consider three questions:

%
\subsubsection{Why do some participants continue to respond to a devalued stimulus?}
During the experiment the security motivational system in all participants generates prospective narratives representing the scenarios in which a shock occurs. From these narratives, the participant infers that a shock will occur if the correct pedal is not pressed when a stimulus appears.  When the electrode is disconnected the narratives are updated, and the participant examines them in order to direct their actions.  Those who responded to the devalued stimulus were unable to form a clear view of the prospective situation at a sub-linguistic level of inference.  If Bayesian learning fails, heuristics may act as a fall back strategy \cite{oreilly2012}, and there is evidence that in a rapidly changing environment, people act according to the last choice that they made \cite{summerfield2011}.  So the behaviour is an attempt to improve recognition with a fallback strategy that is used when Bayesian inference fails.
%
\subsubsection{Why is some participants' understanding not consistent with their behaviour ?}
In the experiment, the participants are aware of the usual conduct of modern psychological experiments, the function of electrical circuits and so on.  These external conditions are coded by narratives at the linguistic level of cognition in both the OCD and control groups.   There are also narratives at a sub-linguistic level using states to represent the stimulus and the shock.  Both levels of narrative contribute to the individual's unitary view of their situation. This global inference remains intact and results in a veridical perspective: the participants who responded to the devalued stimulus were not deluded.   But their behaviour relies on the fast, sub-linguistic level, which is dysfunctional, so participants found their behaviour to be inconsistent with their understanding of the situation.
%
\subsubsection{How do we account for the participants' post-hoc explanations for their behaviour?}
Here we follow the explanation given by Gillan \etalns, who noted that in situations where behaviour contradicts belief, people are known to alter beliefs to match behaviour \cite{gillan2014,festinger1962}.  The need to reconcile beliefs and behaviour is prompted by a request to explain their behaviour but the participants' cognitive model was inconsistent at the time of decision: a sub-linguistic level of inference predicted a shock, while higher levels saw that the electrode was disconnected.  This state of affairs cannot be accurately represented by a simple explanation and it is confusing for the individual who experiences it: they are `in two minds' \cite{evans2003}. So some participants claimed that their action was consistent with their belief while others attributed their behaviour to accident.  

\subsection{Explaining symptoms of OCD} 
OCD covers a broad range of symptoms and there is no set of features both common and peculiar to all subtypes of the disorder \cite[p26]{jakes}.  Rather than confronting the problem of definition, the approach taken here is to explain specific behaviours and symptoms that are found in some individuals with OCD.  This section applies the hierarchical narrative model to checking, fear of harm, intrusive thoughts, and ordering and symmetry obsessions.  
%
\subsubsection{Checking}
We consider an example of compulsive checking where someone repeatedly checks that a gas tap is turned off \cite[p196]{miller1988}.  Our explanation is broadly the same as that for the results of Gillan \etalns's experiment given in the last section.  The individual with OCD has a correct global understanding that the gas tap is off even though there is a failure at a sub-linguistic level of inference.  The individual's SMS relies on this lower level inference so they have to act both to improve the inference and ensure their self-protection.  Since inference about the external world is failing, the individual resorts to their last action under those conditions, and repeatedly checks the gas tap. However, checking does not rectify the failure of recognition, so the SMS does not terminate its response to the perceived threat. Obsessive thoughts, `perhaps the house will burn down' are the inferences made by higher levels of cognition from the counterfactual narratives generated by the failing, sub-linguistic level.
%
\subsubsection{Fear of causing harm}
An example is a case where an individual thinks, without justification, that he has knocked someone down with his car \cite[p196]{miller1988}.  Here and in the case of checking there is a feeling of uncertainty about the true state of affairs -- whether the gas tap is off, and whether an accident has really occurred -- accompanied by an understanding that the obsessional thoughts are a product of his or her own mind \cite{DSM,foa1995}.  
\nind
In individuals with OCD, a failure of inference at a sub-linguistic level of the cognitive hierarchy results in an unclear `view' of the retrospective scene.  The SMS acts both to protect the individual from a potential threat and to correct the failure of inference.  So the individual is forced to finesse their perspective, but since their action does not improve recognition they are compelled to repeat the action. Obsessive thoughts, `perhaps I have run someone over' are the inferences made by higher levels of cognition from the counterfactual narratives generated by the failing, sub-linguistic level.
%
\subsubsection{Intrusive thoughts}
Hudak \etal report a case of intrusive thoughts associated with postpartum OCD \cite{hudak2012}.  A woman was afraid to be alone with her 9 week old son because of terrifying thoughts of stuffing the baby into a microwave oven.  She also reported thoughts about stuffing her husband into a microwave oven which she found even more frightening because she realised that this was physically impossible. 
\nind
In this case, the SMS creates prospective narratives to simulate threat scenarios involving a microwave, but the dysfunction of sub-linguistic inference results in an unclear view of prospective scenes.  So the SMS signals a strong urge to correct the perception in order to ensure that the child is protected.  The SMS usually operates silently, without the individual being aware of the simulated scenarios, but the signal brings the process to the individual's attention.  However, since unitary inference remains intact, the woman `knows' that she is not going to harm her baby, in the sense that she judges this to be the case.  Her judgement is correct and her understanding is veridical, but knowledge is derived from multiple layers, including the failing sub-linguistic layer, so it is partly impaired.  The failure of inference in the sub-linguistic layer has two consequences.  First she has no feeling of knowledge or yedasentience \cite{szechtman2004} and she is compelled to correct this situation.  Second, the failed inference is interpreted by the higher, linguistic layers of cognition and manifests as unusual narratives at a linguistic layer.  There is a further consideration specifically in the case of OCD associated with childbirth: since it is a period of increased vulnerability for the infant there there may also be a natural increase in the generation of unlikely narratives in addition to any dysfunction of inference.  This natural increase may be important in the development of OCD in individuals who are at risk of the disorder.

%
\subsubsection{Ordering and symmetry obsessions}
Here we consider an example of a 38-year-old man who had obsessions with themes of: `(1) the need to know or remember details, (2) the need for exactness in behaviour and precision of expression, and (3) the need for symmetry and sameness in his physical environment (e.g., his appearance, the alignment of books, the condition of belongings). \dots Distress centred on not the content of obsessions, but \dots \, on a tormenting sense of hyperawareness and dissatisfaction.'  \cite{summerfeldt1999}. 
\nind 
The explanations given earlier for checking, fear of harm and intrusive thoughts have implicated a dysfunction of the SMS.  The dysfunction led to a failure to obtain precision in the distribution of $p(N|U)$ at a sub-linguistic level of the cognitive hierarchy. With ordering and symmetry obsessions, the involvement of the SMS is less obvious\footnote{Although some ordering obsessions are associated with a fear of harm if the associated ritual is suppressed\comment{need citation}.}, but the underlying dysfunction affects other functions that rely on modelling external events. Examples of normal ordering behaviour are straightening a painting that is askew or tidying a desk, and these can be understood as active inference, a unified account of action and perception \cite{friston2010,fristonfree}.  In OCD, an imprecise distribution of sub-linguistic narratives leads to ordered scenes appearing `not just right' \cite{coles2003}, and it  results in a compulsion to act.  It is this attempt to achieve recognition by finessing precision that explains not just ordering behaviour but the other manifestations of OCD.
\subsection{Neuropsychological evidence} 
A 2004 review analysed 50 studies of cognitive impairment in OCD and concluded there was no clear and specific neurocognitive profile \cite{kuelz2004}.  However, it found that there was some evidence for visuospatial memory dysfunction in OCD, which they attributed to an underlying executive dysfunction.  They also noted impaired verbal memory for tasks requiring effortful encoding strategies.  A 2013 meta-analysis of 113 studies found an inconsistent picture of deficits, but noted the contrast between a large effect size for verbal memory compared with a small effect size for non-verbal memory \cite{abramovitch2013}. The authors suggested that poorer performance on the  Rey--Osterrieth Complex Figure Test (RCFT) may be related to executive functioning rather than to memory impairments \emph{per se}, in line with a finding by Savage \etal that nonverbal memory problems in OCD subjects were mediated by impaired organisational strategies used during the initial copy of the RCFT figure \etal \cite{savage1999}.  They suggested that the observed memory impairment is secondary to impaired executive strategies used during learning.  These results were not replicated by Shin \etal who found that deficits in organisational strategies mediated non-verbal memory impairment in medicated but not drug-naive patients \cite{shin2010}.  
\nind
Many studies have confounding variables of medication, co-morbidity and heterogeneous groups which include different subtypes, but the impairment of non-verbal memory has been a consistent finding and it contrasts with the relatively lesser impairment of verbal memory \cite{muller2005}.  Further, \citet{cuttler2007} found that prospective memory was impaired in a sub-clinical group of compulsive checkers.  The group performed poorly at remembering to perform a task in response to an event, though not when having to remember to do something at some future time.  This impairment of event-based, but not time-based prospective memory was replicated using a sample of clinical OCD checkers \cite{harris2010}.  Additionally, \citet{harris2010} found that clinical OCD checkers, unlike the sub-clinical group, had no more complaints about their perceived prospective memory ability than controls.
\nind
Jaafari \etal applied reading span and backward location span tests to a group of 32 OCD patients and a group of individually matched controls \cite{jaafari2013}.  They also examined checking behaviour by using an image comparison test with eye-tracking equipment to count gaze moves between the images.  They found that both the verbal and visuospatial components of working memory \comment{baddeley1992}were impaired in OCD participants, and that the OCD group made more gaze moves to compare images than the controls.  Further, the patients' deficit in the comparison task was negatively correlated with their working memory spans.  There was no dependence of the working memory or gaze move scores on medication status.
\subsubsection{Interpretation by predictive coding}
How might we interpret these results in terms of predictive coding?  According to a generative model, the perceptual apparatus generates an output to minimise error with incoming sense data, as illustrated in Figure \ref{fig:perception}.  A severe dysfunction in the generative process can lead to the visual hallucinations of Charles Bonnet Syndrome, whereas in this case the dysfunction leads to impaired recognition.  In Jaafari \etalns's experiment \cite{jaafari2013} the participant first views one image, then a second image on the same screen to make their comparison.  During the comparison the participant recalls the first image by generating sub-linguistic narratives to model the image constituents.  The participant compares the sensory data from the second image with these narratives to determine if the images are the same.  In OCD the distribution of narratives generated for the first image is abnormally broad, as illustrated in Figure \ref{fig:dysfn}, so the outcome of the comparison process is uncertain.  In consequence the participant looks back to the original image to finesse recognition, in other words to refresh their memory.   The increased number of gaze moves made by the OCD group compared with controls is a manifestation of impaired recognition. We can compare this recognition failure with uncorrected vision: just as an optical defect leads to blurred vision, a generative dysfunction leads to uncertain inference. 
\nind
We see a similar failure of recognition in Cuttler and Graf's finding that event-based but not time-based prospective memory was impaired \cite{cuttler2007} in a group sub-clinical compulsive checkers.  To test event-based memory, they asked participants to give them a personal belonging at the start of the experiment, and told them to ask for it back in response to a specific verbal cue: `we are now finished with all of the tests'.  At the end of the experiment, the experimenter gave the cue and allowed 10 seconds for the participant to respond.  Among the 126 sub-clinical checkers tested, 56\% of high checkers responded compared with 80\% of low checkers.  Again we interpret these results in terms of a failure of recognition at a sub-linguistic level leading to a failure to recognise the cue. As with Jaafari \etalns's experiment \cite{jaafari2013} the participant compares sensory data, in this case an auditory cue, with narratives generated from their internal model.  The broad distribution of generated narratives impairs the comparison process, so that the participant fails to recognise the cue.  In contrast the time-based prospective memory remains unimpaired because it depends only on the individual monitoring the time to make their response.


%
\subsection{Testing the hypothesis} 
We claim that in OCD recognition at a sub-linguistic level is impaired while global inference is intact.  Two testable predictions can be derived from this claim.  First, it should be possible to detect the failure of recognition at a sub-linguistic level of inference. Second, individuals with OCD should have an enhanced ability to make correct inferences in the presence of uncertainty and errors.  There are some challenges in testing these predictions and we offer just an outline for what might be tested, and suggest some experimental paradigms.
\nind
The design of behavioural experiments must allow for the correction or errors in sub-linguistic inference by global inference. Stokes \etal gave an example of probing cognitive variables using EEG, rather than relying only on a task design \cite{stokes2014}.  Using patterned images as stimuli, they used EEG and reaction time measurements to probe learning of cues indicating that a target stimulus is likely to follow.   Participants were instructed to press a button as quickly as possible each time they detected the target image.  A possible variation of their experiment is include some cue images that are similar to but distinct from the target cues, perhaps by transforming the original cue images (using rotation, mirror etc.).  Participants with OCD should respond more to these pseudo-cues than would healthy individuals.  Another possible paradigm is that employed by Zhao \etal who found that ordered triples of shapes bias attention towards their location, compared with random sequences \cite{zhao2013}.  Again, the experiment could be modified to include similar but distinct images in the triples: people with OCD would fail to distinguish these from the real cues.  Since the recognition errors might be a function of a threat or reward context, it is necessary to test dependence on these variables.
\nind
To test enhanced global inference in the presence of errors, we could measure reading speed and comprehension on a piece of text such as the example in section \ref{sec:dys}.  The deficit relative to the score without errors would indicate their robustness of global inference to error.  But reading style depends on education and previous learning, so a similar experiment using images or patterns may have fewer confounds.  Another approach is to test the participant's ability to make correct inferences in a complex, underdetermined situation: people with OCD should perform better.  One approach is to probe inferences from text where words have been deleted, so leaving the text open to multiple interpretations.  A multiple-choice questionnaire would then allow the participant to select the most likely inference.  People with OCD are, generally, known to show poorer performance relative to healthy individuals across most neuropsychological domains \cite{abramovitch2013}.  So in testing for enhanced inference ability, we would need to limit the group those who show checking behaviours, are unmedicated and have no known comorbidity.  

%

\section{Discussion}
Our model consists of the hierarchical narrative framework of cognition and the dysfunction which is hypothesised to cause OCD.   We refer to this model of OCD as the `hierarchical narrative model' (HNM).  Here we consider how it compares with alternative models for OCD in terms of fit to known data.  
\nind
We begin by comparing the HNM with the habit hypothesis that is supported by Gillan \etalns's experimental results.  The explanation of compulsions as habits does little to elucidate the mechanism: it does not go far beyond a description of the behaviour.  The HNM explains the behaviour as an attempt to improve recognition with a fallback strategy which is used when Bayesian inference fails.  This fallback is known to occur when information is changing too rapidly for the cognitive system, so it has support from existing data \cite{oreilly2012,summerfield2011}.  Gillan \etal explain obsessional thoughts as a means of resolving the dissonance between behaviour and belief.  Again, the HNM goes further by explaining this dissonance as a mismatch between global and sub-linguistic inference.  So the model does not conflict with the habit hypothesis but instead subsumes and extends it.
\nind
The cognitive-behavioural model of OCD sees obsessions as a universal phenomenon which turn into abnormal obsessions when negatively appraised, for example as implying some personal responsibility for the thought's content \cite{salkovskis1985}.  This account was criticised by Jakes \cite[p44]{jakes} on grounds of parsimony: if negative appraisals result in abnormal obsessions, what causes the `normal' obsessions that occur in the general population?   In the HNM they arise from from the normal operation of a security mechanism which generates counterfactuals.  Obsessions in OCD occur by the same mechanism, except that in OCD the unlikely scenarios occur more frequently. So the HNM explains normal and abnormal obsessions using a single cognitive model, and explains abnormal obsessions with a hypothesised dysfunction.  Further, the cognitive-behavioural model does not explain the strange content of some obsessions in OCD \cite{rassin2007}. The HNM explains these obsessions as the interpretation by the higher, linguistic layers of cognition of the unusual narratives from the sub-linguistic layer.  So, unlike the cognitive-behavioural model, the HNM elucidates the mechanism by which obsessions occur and explains how they differ from normal obsessions. 
\nind
According to the cognitive-behavioural model compulsions are understood as normal responses triggered by irrational beliefs  \cite{salkovskis1985}\cite{rachman1997}.  The negative appraisal of an obsession leads to an the individual `neutralising' the obsession, as when a mother suffering from postpartum OCD avoids her baby in order to prevent harm.  However, this account does not explain why individuals with OCD recognise that the obsessional thoughts are a product of his or her own mind \cite{DSM,foa1995}.  The HNM explains this phenomenon using a dual process theory.  That is, the obsessions have a source at a sub-linguistic level of inference, while global inference - the unitary perception of the individual - is unimpaired.  Compulsions are understood as cognitive system automatically finessing inference to preserve their knowledge of the environment. This account explains both the strength of the compulsions, and their irrational and ego-dystonic nature.  
\nind
\subsection{Neuropsychological models}
Some theories of OCD have implicated problems with memory or worry about a failure of prospective \cite{muller2005,cuttler2007}.   \citet{harris2010} found that a clinical OCD group had no more concern about their prospective memory than a control group, and so amended this cause as explaining the \emph{development} rather than the maintenance of checking compulsions. Yet they also found that both non-clinical and clinical checkers show a deficit in performance on an event-based memory task.  This failure in memory is also observed in the results from \citet{jaafari2013,abramovitch2013}.  The HNM predicts problems with memory because there is a failure of recognition which impedes coding, so the model subsumes and extends the memory hypothesis.
\nind
Where the HNM has less to contribute is in interpretation of imaging data.  Here the the orbitofronto-striatal model and its variants support the data \cite{menzies2008}.  However, the future development of the HNM, for example by specifying the kind of generative model, may allow for more precise predictions.   In particular the default mode network is thought to be engaged by self-projection, including thinking about the future, episodic remembering and theory of mind \cite{buckner2007,spreng2010}.  Abnormalities of the default mode network have been found in bipolar disorder and schizophrenia \cite{ongur2010}, and there is increased interest in its implications for OCD \cite{stern2012,beucke2014}.  If the hypothesised dysfunction could be identified with the observed abnormalities, this would be strong evidence for the model.

%

\section{Conclusion}
We have presented a model for OCD using a predictive coding framework and a hypothesised dysfunction.  The framework includes the concept of a formal narrative which follows from the use of generative models for inference. It consists of the following elements,
\vspace{-3mm}\paragraph{1)} A hierarchical narrative model of cognition has a hierarchy of inference layers, each of which contributes to the individual's global view of their environment.  


\vspace{-3mm} \paragraph{2)} Narratives are constantly generated to model the past, present and future in response to currently inferred states.  These narratives are evaluated for risk and reward.

\vspace{-3mm}\paragraph{}
We suggested that an illustrative metaphor for the hierarchical narrative model could be a multidisciplinary clinical team.  Each team member forms a view from their prior beliefs and the evidence they see, and the global consensus is formed from the views of all the members.  We proposed that OCD arises from a dysfunction at a sub-linguistic level of cognition while global inference remains broadly unimpaired. The consequent failure of recognition is experienced as the external world being `not just right', and its automatic correction is felt as compulsion.  The model explains a wide variety of symptoms and some neuropsychological findings.  We outlined some experimental work to test the predictions made from hypothesis, and so to establish its association with OCD.  Finally, we compared the model with existing theoretical approaches to OCD in terms of parsimony and fit to existing data.  We conclude that the model provides a deeper explanation for behavioural observations than current models, and that it has potential for further development for application to neuropsychological data.
\newpage
\bibliographystyle{apalike}
\bibliography{narr}
\end{document}

%% file: macpap2.tex
\newcommand{\be}{\begin{equation}}
\newcommand{\ee}{\end{equation}}
\newcommand{\bean}{\begin{eqnarray}}
\newcommand{\eean}{\end{eqnarray}}
\newcommand{\bea}{\begin{eqnarray*}}
\newcommand{\eea}{\end{eqnarray*}}
\newcommand{\bfig}{\begin{figure}}
\newcommand{\efig}{\end{figure}}
\newcommand{\ba}{\begin{array}}
\newcommand{\ea}{\end{array}}

\newcommand{\nind}{\newline\indent}

\newcommand{\comment}[1]{}

\newcommand{\etal}{\emph{et al. }}
\newcommand{\etalns}{\emph{et al.}}


%% file: narr.bbl
\begin{thebibliography}{}

\bibitem[Abramovitch et~al., 2013]{abramovitch2013}
Abramovitch, A., Abramowitz, J.~S., and Mittelman, A. (2013).
\newblock The neuropsychology of adult obsessive--compulsive disorder: A
  meta-analysis.
\newblock {\em Clinical psychology review}, 33(8):1163--1171.

\bibitem[Abramowitz et~al., 2006]{abramowitz2006}
Abramowitz, J.~S., Khandker, M., Nelson, C.~A., Deacon, B.~J., and Rygwall, R.
  (2006).
\newblock The role of cognitive factors in the pathogenesis of
  obsessive--compulsive symptoms: A prospective study.
\newblock {\em Behaviour Research and Therapy}, 44(9):1361--1374.

\bibitem[Abramowitz et~al., 2007]{abramowitz2007}
Abramowitz, J.~S., Nelson, C.~A., Rygwall, R., and Khandker, M. (2007).
\newblock The cognitive mediation of obsessive-compulsive symptoms: A
  longitudinal study.
\newblock {\em Journal of Anxiety Disorders}, 21(1):91--104.

\bibitem[Abramowitz et~al., 2009]{abramowitz2009}
Abramowitz, J.~S., Taylor, S., and McKay, D. (2009).
\newblock Obsessive-compulsive disorder.
\newblock {\em The Lancet}, 374(9688):491 -- 499.

\bibitem[Antony and Barlow, 2002]{antony02}
Antony, M.~M. and Barlow, D.~H. (2002).
\newblock {\em Handbook of assessment and treatment planning for psychological
  disorders}.
\newblock Guilford press.

\bibitem[Bayne, 2009]{bayneschol}
Bayne, T. (2009).
\newblock {U}nity of consciousness.
\newblock {\em Scholarpedia}, 4(2):7414.
\newblock {revision 52128}.

\bibitem[Bayne, 2010]{bayne2010}
Bayne, T. (2010).
\newblock {\em The unity of consciousness}.
\newblock Oxford University Press.

\bibitem[Beucke et~al., 2014]{beucke2014}
Beucke, J.~C., Sepulcre, J., Eldaief, M.~C., Sebold, M., Kathmann, N., and
  Kaufmann, C. (2014).
\newblock Default mode network subsystem alterations in obsessive-compulsive
  disorder.
\newblock {\em The British Journal of Psychiatry}, 205(5):376--382.

\bibitem[Buckner and Carroll, 2007]{buckner2007}
Buckner, R.~L. and Carroll, D.~C. (2007).
\newblock Self-projection and the brain.
\newblock {\em Trends in cognitive sciences}, 11(2):49--57.

\bibitem[Coles et~al., 2003]{coles2003}
Coles, M.~E., Frost, R.~O., Heimberg, R.~G., and Rh{\'e}aume, J. (2003).
\newblock “{N}ot just right experiences”: perfectionism,
  obsessive--compulsive features and general psychopathology.
\newblock {\em Behaviour Research and Therapy}, 41(6):681--700.

\bibitem[Corlett et~al., 2010]{corlett}
Corlett, P., Taylor, J., Wang, X.-J., Fletcher, P., and Krystal, J. (2010).
\newblock Toward a neurobiology of delusions.
\newblock {\em Progress in neurobiology}, 92(3):345--369.

\bibitem[Cougle and Lee, 2014]{cougle2014}
Cougle, J.~R. and Lee, H.-J. (2014).
\newblock Pathological and non-pathological features of obsessive-compulsive
  disorder: Revisiting basic assumptions of cognitive models.
\newblock {\em Journal of Obsessive-Compulsive and Related Disorders},
  3(1):12--20.

\bibitem[Cuttler and Graf, 2007]{cuttler2007}
Cuttler, C. and Graf, P. (2007).
\newblock Sub-clinical compulsive checkers’ prospective memory is impaired.
\newblock {\em Journal of Anxiety Disorders}, 21(3):338--352.

\bibitem[DSM, 2000]{DSM}
DSM (2000).
\newblock {\em Diagnostic and statistical manual of mental disorders:
  DSM-IV-TR.}
\newblock American Psychiatric Publishing, Inc, Washington, DC.

\bibitem[Evans, 2003]{evans2003}
Evans, J. S.~B. (2003).
\newblock In two minds: dual-process accounts of reasoning.
\newblock {\em Trends in cognitive sciences}, 7(10):454--459.

\bibitem[Festinger, 1962]{festinger1962}
Festinger, L. (1962).
\newblock {\em A theory of cognitive dissonance}, volume~2.
\newblock Stanford {U}niversity press.

\bibitem[Fletcher and Frith, 2008]{fletcher}
Fletcher, P.~C. and Frith, C.~D. (2008).
\newblock Perceiving is believing: a {B}ayesian approach to explaining the
  positive symptoms of schizophrenia.
\newblock {\em Nature Reviews Neuroscience}, 10(1):48--58.

\bibitem[Foa and Kozak, 1995]{foa1995}
Foa, E.~B. and Kozak, M.~J. (1995).
\newblock {DSM-IV} field trial: obsessive-compulsive disorder.
\newblock {\em The American journal of psychiatry}.

\bibitem[Frankish and Ramsey, 2012]{frankish2012}
Frankish, K. and Ramsey, W. (2012).
\newblock {\em The {C}ambridge handbook of cognitive science}.
\newblock Cambridge University Press.

\bibitem[Friston, 2005a]{fristoncort}
Friston, K. (2005a).
\newblock A theory of cortical responses.
\newblock {\em Philosophical Transactions of the Royal Society B: Biological
  Sciences}, 360(1456):815--836.

\bibitem[Friston, 2010]{fristonfree}
Friston, K. (2010).
\newblock The free-energy principle: a unified brain theory?
\newblock {\em Nature Reviews Neuroscience}, 11(2):127--138.

\bibitem[Friston, 2012]{fristonhist}
Friston, K. (2012).
\newblock The history of the future of the {B}ayesian brain.
\newblock {\em NeuroImage}, 62(2):1230--1233.

\bibitem[Friston et~al., 2011]{friston2011}
Friston, K., Mattout, J., and Kilner, J. (2011).
\newblock Action understanding and active inference.
\newblock {\em Biological cybernetics}, 104(1-2):137--160.

\bibitem[Friston, 2005b]{fristonhal}
Friston, K.~J. (2005b).
\newblock Hallucinations and perceptual inference.
\newblock {\em Behavioral and Brain Sciences}, 28(06):764--766.

\bibitem[Friston et~al., 2010]{friston2010}
Friston, K.~J., Daunizeau, J., Kilner, J., and Kiebel, S.~J. (2010).
\newblock Action and behavior: a free-energy formulation.
\newblock {\em Biological cybernetics}, 102(3):227--260.

\bibitem[Gibbs, 1996]{gibbs1996}
Gibbs, N.~A. (1996).
\newblock Nonclinical populations in research on obsessive-compulsive disorder:
  A critical review.
\newblock {\em Clinical Psychology Review}, 16(8):729--773.

\bibitem[Gillan et~al., 2014]{gillan2014}
Gillan, C.~M., Morein-Zamir, S., Urcelay, G.~P., Sule, A., Voon, V.,
  Apergis-Schoute, A.~M., Fineberg, N.~A., Sahakian, B.~J., and Robbins, T.~W.
  (2014).
\newblock Enhanced avoidance habits in obsessive-compulsive disorder.
\newblock {\em Biological psychiatry}, (75):631--638.

\bibitem[Graybiel and Rauch, 2000]{graybiel2000}
Graybiel, A. and Rauch, S. (2000).
\newblock Toward a neurobiology of obsessive-compulsive disorder.
\newblock {\em Neuron}, 28(2):343--347.

\bibitem[Harris et~al., 2010]{harris2010}
Harris, L.~M., Vaccaro, L., Jones, M.~K., and Boots, G.~M. (2010).
\newblock Evidence of impaired event-based prospective memory in clinical
  obsessive--compulsive checking.
\newblock {\em Behaviour Change}, 27(02):84--92.

\bibitem[Hohwy, 2013]{hohwy}
Hohwy, J. (2013).
\newblock {\em The predictive mind}.
\newblock Oxford University Press.

\bibitem[Hudak and Wisner, 2012]{hudak2012}
Hudak, R. and Wisner, K.~L. (2012).
\newblock Diagnosis and treatment of postpartum obsessions and compulsions that
  involve infant harm.
\newblock {\em American Journal of Psychiatry}, 169(4):360--363.

\bibitem[Jaafari et~al., 2013]{jaafari2013}
Jaafari, N., Frasca, M., Rigalleau, F., Rachid, F., Gil, R., Oli{\'e}, J.-P.,
  Guehl, D., Burbaud, P., Aouizerate, B., Rotg{\'e}, J.-Y., et~al. (2013).
\newblock Forgetting what you have checked: A link between working memory
  impairment and checking behaviors in obsessive-compulsive disorder.
\newblock {\em European Psychiatry}, 28(2):87--93.

\bibitem[Jakes, 1989a]{jakes1989a}
Jakes, I. (1989a).
\newblock Salkovskis on obsessional-compulsive neurosis: a critique.
\newblock {\em Behaviour research and therapy}, 27(6):673--675.

\bibitem[Jakes, 1989b]{jakes1989b}
Jakes, I. (1989b).
\newblock Salkovskis on obsessional-compulsive neurosis: a rejoinder.
\newblock {\em Behaviour research and therapy}, 27(6):683--684.

\bibitem[Jakes, 2006]{jakes}
Jakes, I. (2006).
\newblock {\em Theoretical approaches to obsessive-compulsive disorder}.
\newblock Cambridge University Press.

\bibitem[Julien et~al., 2007]{julien2007}
Julien, D., O'Connor, K.~P., and Aardema, F. (2007).
\newblock Intrusive thoughts, obsessions, and appraisals in
  obsessive--compulsive disorder: A critical review.
\newblock {\em Clinical psychology review}, 27(3):366--383.

\bibitem[Knill and Pouget, 2004]{knill}
Knill, D.~C. and Pouget, A. (2004).
\newblock The {B}ayesian brain: the role of uncertainty in neural coding and
  computation.
\newblock {\em TRENDS in Neurosciences}, 27(12):712--719.

\bibitem[Kuelz et~al., 2004]{kuelz2004}
Kuelz, A.~K., Hohagen, F., and Voderholzer, U. (2004).
\newblock Neuropsychological performance in obsessive-compulsive disorder: a
  critical review.
\newblock {\em Biological psychology}, 65(3):185--236.

\bibitem[Leckman et~al., 1997]{leckman97}
Leckman, J.~F., Grice, D.~E., Boardman, J., Zhang, H., Vitale, A., Bondi, C.,
  Alsobrook, J., Peterson, B.~S., Cohen, D.~J., Rasmussen, S.~A., et~al.
  (1997).
\newblock Symptoms of obsessive-compulsive disorder.
\newblock {\em American Journal of Psychiatry}, 154(7):911--917.

\bibitem[Leonard and Swedo, 2001]{leonard2001}
Leonard, H.~L. and Swedo, S.~E. (2001).
\newblock Paediatric autoimmune neuropsychiatric disorders associated with
  streptococcal infection {(PANDAS)}.
\newblock {\em The International Journal of Neuropsychopharmacology},
  4(02):191--198.

\bibitem[McFall and Wollersheim, 1979]{mcfall1979}
McFall, M.~E. and Wollersheim, J.~P. (1979).
\newblock Obsessive-compulsive neurosis: A cognitive-behavioral formulation and
  approach to treatment.
\newblock {\em Cognitive Therapy and Research}, 3(4):333--348.

\bibitem[McKay et~al., 2004]{mckay2004}
McKay, D., Abramowitz, J.~S., Calamari, J.~E., Kyrios, M., Radomsky, A.,
  Sookman, D., Taylor, S., and Wilhelm, S. (2004).
\newblock A critical evaluation of obsessive--compulsive disorder subtypes:
  symptoms versus mechanisms.
\newblock {\em Clinical Psychology Review}, 24(3):283--313.

\bibitem[Menzies et~al., 2008]{menzies2008}
Menzies, L., Chamberlain, S.~R., Laird, A.~R., Thelen, S.~M., Sahakian, B.~J.,
  and Bullmore, E.~T. (2008).
\newblock Integrating evidence from neuroimaging and neuropsychological studies
  of obsessive-compulsive disorder: the orbitofronto-striatal model revisited.
\newblock {\em Neuroscience \& Biobehavioral Reviews}, 32(3):525--549.

\bibitem[Miller and Cooper, 1988]{miller1988}
Miller, E. and Cooper, P.~J. (1988).
\newblock {\em Adult abnormal psychology}.
\newblock Churchill Livingstone.

\bibitem[Mitchell, 2009]{mitchell2009}
Mitchell, J.~P. (2009).
\newblock Inferences about mental states.
\newblock {\em Philosophical Transactions of the Royal Society B: Biological
  Sciences}, 364(1521):1309--1316.

\bibitem[Moore, 2014]{thesis}
Moore, P.~J. (2014).
\newblock {\em Mathematical Modelling, Forecasting and Telemonitoring of Mood
  in Bipolar Disorder}.
\newblock PhD thesis, Oxford University.

\bibitem[Muller and Roberts, 2005]{muller2005}
Muller, J. and Roberts, J.~E. (2005).
\newblock Memory and attention in obsessive--compulsive disorder: a review.
\newblock {\em Journal of anxiety disorders}, 19(1):1--28.

\bibitem[OCCWG, 2003]{occwg03_1}
OCCWG (2003).
\newblock Psychometric validation of the obsessive beliefs questionnaire and
  the interpretation of intrusions inventory: Part 1.
\newblock {\em Behaviour Research and Therapy}, 41(8):863.

\bibitem[OCCWG, 2005]{occwg03_2}
OCCWG (2005).
\newblock Psychometric validation of the obsessive belief questionnaire and
  interpretation of intrusions inventory: Part 2: Factor analyses and testing
  of a brief version.
\newblock {\em Behaviour Research and Therapy}, 43(11):1527--1542.

\bibitem[{\"O}ng{\"u}r et~al., 2010]{ongur2010}
{\"O}ng{\"u}r, D., Lundy, M., Greenhouse, I., Shinn, A.~K., Menon, V., Cohen,
  B.~M., and Renshaw, P.~F. (2010).
\newblock Default mode network abnormalities in bipolar disorder and
  schizophrenia.
\newblock {\em Psychiatry Research: Neuroimaging}, 183(1):59--68.

\bibitem[O{'}Reilly et~al., 2012]{oreilly2012}
O{'}Reilly, J.~X., Jbabdi, S., and Behrens, T.~E. (2012).
\newblock How can a {B}ayesian approach inform neuroscience?
\newblock {\em European Journal of Neuroscience}, 35(7):1169--1179.

\bibitem[Rachman, 1993]{rachman1993}
Rachman, S. (1993).
\newblock Obsessions, responsibility and guilt.
\newblock {\em Behaviour research and therapy}, 31(2):149--154.

\bibitem[Rachman, 1997]{rachman1997}
Rachman, S. (1997).
\newblock A cognitive theory of obsessions.
\newblock {\em Behaviour Research and Therapy}, 35(9):793--802.

\bibitem[Rachman and de~Silva, 1978]{rachman1978}
Rachman, S. and de~Silva, P. (1978).
\newblock Abnormal and normal obsessions.
\newblock {\em Behaviour research and therapy}, 16(4):233--248.

\bibitem[Rapoport, 1990]{rapoport1990}
Rapoport, J.~L. (1990).
\newblock Obsessive compulsive disorder and basal ganglia dysfunction.
\newblock {\em Psychological medicine}, 20(03):465--469.

\bibitem[Rassin et~al., 2007]{rassin2007}
Rassin, E., Cougle, J.~R., and Muris, P. (2007).
\newblock Content difference between normal and abnormal obsessions.
\newblock {\em Behaviour Research and Therapy}, 45(11):2800--2803.

\bibitem[Reichert et~al., 2013]{reichert2013}
Reichert, D.~P., Seri{\`e}s, P., and Storkey, A.~J. (2013).
\newblock {Charles Bonnet syndrome: evidence for a generative model in the
  cortex?}
\newblock {\em PLoS computational biology}, 9(7):e1003134.

\bibitem[Salkovskis, 1985]{salkovskis1985}
Salkovskis, P.~M. (1985).
\newblock Obsessional-compulsive problems: A cognitive-behavioural analysis.
\newblock {\em Behaviour research and therapy}, 23(5):571--583.

\bibitem[Salkovskis, 1989]{salkovskis1989}
Salkovskis, P.~M. (1989).
\newblock Cognitive-behavioural factors and the persistence of intrusive
  thoughts in obsessional problems.
\newblock {\em Behaviour research and therapy}, 27(6):677--682.

\bibitem[Savage et~al., 1999]{savage1999}
Savage, C.~R., Baer, L., Keuthen, N.~J., Brown, H.~D., Rauch, S.~L., and
  Jenike, M.~A. (1999).
\newblock Organizational strategies mediate nonverbal memory impairment in
  obsessive--compulsive disorder.
\newblock {\em Biological psychiatry}, 45(7):905--916.

\bibitem[Saxena et~al., 2001]{saxena2001}
Saxena, S., Bota, R.~G., and Brody, A.~L. (2001).
\newblock Brain-behavior relationships in obsessive-compulsive disorder.
\newblock In {\em Seminars in clinical neuropsychiatry}, volume~6, pages
  82--101.

\bibitem[Shafran and Rachman, 2004]{shafran2004}
Shafran, R. and Rachman, S. (2004).
\newblock Thought-action fusion: a review.
\newblock {\em Journal of Behavior Therapy and Experimental Psychiatry},
  35(2):87--107.

\bibitem[Shannon, 1948]{shannon}
Shannon, C.~E. (1948).
\newblock A mathematical theory of communication.
\newblock {\em The Bell System Technical Journal}, 27:379--423, 623--656.

\bibitem[Shin et~al., 2010]{shin2010}
Shin, N.~Y., Kang, D.-H., Choi, J.-S., Jung, M.~H., Jang, J.~H., and Kwon,
  J.~S. (2010).
\newblock Do organizational strategies mediate nonverbal memory impairment in
  drug-na{\"\i}ve patients with obsessive-compulsive disorder?
\newblock {\em Neuropsychology}, 24(4):527.

\bibitem[Southall and von Helmholtz, 1925]{southall1925}
Southall, J. P.~C. and von Helmholtz, H. (1925).
\newblock {\em Helmholtz's treatise on physiological optics}.
\newblock Optical Society of America.

\bibitem[Spreng and Grady, 2010]{spreng2010}
Spreng, R.~N. and Grady, C.~L. (2010).
\newblock Patterns of brain activity supporting autobiographical memory,
  prospection, and theory of mind, and their relationship to the default mode
  network.
\newblock {\em Journal of cognitive neuroscience}, 22(6):1112--1123.

\bibitem[Stern et~al., 2012]{stern2012}
Stern, E.~R., Fitzgerald, K.~D., Welsh, R.~C., Abelson, J.~L., and Taylor,
  S.~F. (2012).
\newblock Resting-state functional connectivity between fronto-parietal and
  default mode networks in obsessive-compulsive disorder.
\newblock {\em PloS one}, 7(5):e36356.

\bibitem[Stokes et~al., 2014]{stokes2014}
Stokes, M.~G., Myers, N.~E., Turnbull, J., and Nobre, A.~C. (2014).
\newblock Preferential encoding of behaviorally relevant predictions revealed
  by {EEG}.
\newblock {\em Frontiers in human neuroscience}, 8.

\bibitem[Suddendorf and Corballis, 1997]{suddendorf1997}
Suddendorf, T. and Corballis, M.~C. (1997).
\newblock Mental time travel and the evolution of the human mind.
\newblock {\em Genetic, social, and general psychology monographs},
  123(2):133--167.

\bibitem[Summerfeldt, 2004]{summerfeldt2004}
Summerfeldt, L.~J. (2004).
\newblock Understanding and treating incompleteness in obsessive-compulsive
  disorder.
\newblock {\em Journal of clinical psychology}, 60(11):1155--1168.

\bibitem[Summerfeldt et~al., 1999]{summerfeldt1999}
Summerfeldt, L.~J., Richter, M.~A., Antony, M.~M., and Swinson, R.~P. (1999).
\newblock Symptom structure in obsessive-compulsive disorder: a confirmatory
  factor-analytic study.
\newblock {\em Behaviour Research and Therapy}, 37(4):297--311.

\bibitem[Summerfield et~al., 2011]{summerfield2011}
Summerfield, C., Behrens, T.~E., and Koechlin, E. (2011).
\newblock Perceptual classification in a rapidly changing environment.
\newblock {\em Neuron}, 71(4):725--736.

\bibitem[Swedo, 2002]{swedo2002}
Swedo, S.~E. (2002).
\newblock Pediatric autoimmune neuropsychiatric disorders associated with
  streptococcal infections {(PANDAS)}.
\newblock {\em Molecular Psychiatry}, 7:S24--S25.

\bibitem[Szechtman and Woody, 2004]{szechtman2004}
Szechtman, H. and Woody, E. (2004).
\newblock Obsessive-compulsive disorder as a disturbance of security
  motivation.
\newblock {\em Psychological review}, 111(1):111.

\bibitem[Taylor et~al., 2006]{taylor06}
Taylor, S., Abramowitz, J.~S., McKay, D., Calamari, J.~E., Sookman, D., Kyrios,
  M., Wilhelm, S., and Carmin, C. (2006).
\newblock Do dysfunctional beliefs play a role in all types of
  obsessive--compulsive disorder?
\newblock {\em Journal of Anxiety Disorders}, 20(1):85--97.

\bibitem[Taylor et~al., 2005]{taylor2005}
Taylor, S., McKay, D., and Abramowitz, J.~S. (2005).
\newblock Is obsessive-compulsive disorder a disturbance of security
  motivation? comment on {S}zechtman and {W}oody (2004).

\bibitem[Von~Helmholtz, 1866]{helm}
Von~Helmholtz, H. (1866).
\newblock {\em Handbuch der physiologischen Optik: mit 213 in den Text
  eingedruckten Holzschnitten und 11 Tafeln}, volume~9.
\newblock Voss.

\bibitem[Whiteside et~al., 2004]{whiteside2004}
Whiteside, S.~P., Port, J.~D., and Abramowitz, J.~S. (2004).
\newblock A meta--analysis of functional neuroimaging in obsessive--compulsive
  disorder.
\newblock {\em Psychiatry Research: Neuroimaging}, 132(1):69--79.

\bibitem[Wise and Rapoport, 1989]{wise1989}
Wise, S.~P. and Rapoport, J.~L. (1989).
\newblock Obsessive-compulsive disorder: is it basal ganglia dysfunction.
\newblock {\em Obsessive-compulsive disorder in children and adolescents},
  pages 327--44.

\bibitem[Woody and Szechtman, 2005]{woody2005}
Woody, E.~Z. and Szechtman, H. (2005).
\newblock Motivation, time course, and heterogeneity in obsessive-compulsive
  disorder: {Response to Taylor, McKay, and Abramowitz} (2005).

\bibitem[Zhao et~al., 2013]{zhao2013}
Zhao, J., Al-Aidroos, N., and Turk-Browne, N.~B. (2013).
\newblock Attention is spontaneously biased toward regularities.
\newblock {\em Psychological Science}, 24.

\end{thebibliography}
